\documentclass[12pt]{article}
\pdfoutput=1

\usepackage[utf8]{inputenc}
\usepackage[left=2.55cm, right=2.55cm, top=2.55cm, bottom=2.55cm]{geometry}
\usepackage{amsmath,amssymb,amsbsy}
\usepackage{slashed}
\usepackage{xcolor}
\usepackage{graphicx}
\usepackage{url}
\usepackage{cancel}
\usepackage{cite}
\usepackage[colorlinks=true,allcolors=blue,pdfborder={0 0 0},linktocpage=false,pdfencoding=auto]{hyperref}
\usepackage{tabularx,booktabs}
\usepackage{multicol}
\usepackage{feynmp}
\usepackage{units}
\usepackage{xspace}
\usepackage[labelfont=bf]{caption}
\usepackage[section]{placeins}
\usepackage{subcaption}
\usepackage{soul} 
\usepackage{bbold}
\usepackage{parskip}
\usepackage{float}

\definecolor{darkred}{rgb}{0.6,0,0}
\definecolor{darkpurple}{rgb}{0.5,0,0.5}

\newcommand\snowmass{\begin{center}\rule[-0.2in]{\hsize}{0.01in}\\\rule{\hsize}{0.01in}\\
\vskip 0.1in Submitted to the  Proceedings of the US Community Study\\ 
on the Future of Particle Physics (Snowmass 2021)\\ 
\rule{\hsize}{0.01in}\\\rule[+0.2in]{\hsize}{0.01in} \end{center}}

\def\B0{B^{(0)}}
\def\A0{A_3^{(0)}}

\begin{document}

\author{Amin Aboubrahim$^a$\footnote{\href{mailto:aabouibr@uni-muenster.de}{aabouibr@uni-muenster.de}}~, Wan-Zhe Feng$^b$\footnote{\href{mailto:vicf@tju.edu.cn}{vicf@tju.edu.cn}}~, Pran Nath$^c$\footnote{\href{mailto:p.nath@northeastern.edu}{p.nath@northeastern.edu}}~ and Zhu-Yao Wang$^c$\footnote{\href{mailto:wang.zhu@northeastern.edu}{wang.zhu@northeastern.edu}} \\~\\
$^{a}$\textit{\normalsize Institut f\"ur Theoretische Physik, Westf\"alische Wilhelms-Universit\"at M\"unster,} \\
\textit{\normalsize Wilhelm-Klemm-Stra{\ss}e 9, 48149 M\"unster, Germany} \\
$^{b}$\textit{\normalsize Center for Joint Quantum Studies and Department of Physics,}\\
\textit{\normalsize School of Science, Tianjin University, Tianjin 300350, PR. China}\\
$^{c}$\textit{\normalsize Department of Physics, Northeastern University,
Boston, MA 02115-5000, USA} \\}

\title{\vspace{-3cm}\begin{flushright}
{\small MS-TP-21-24}
\end{flushright}
\vspace{0.5cm}
\Large{ \bf Hidden sectors and a multi-temperature universe}
\vspace{0.5cm}}

\date{}
\maketitle

\vspace{-1cm}
\snowmass

\begin{abstract}

A variety of supergravity and string based models contain hidden sectors which can play a role
in particle physics phenomena and in cosmology.
 In this note we discuss the possibility
that the visible sector and the hidden sectors in general live in different heat baths.
 Further, it is
entirely possible that dark matter resides partially or wholly  in hidden sectors in the form of dark Dirac
 fermions, dark neutralinos or dark photons.  A proper analysis of dark matter and of dark forces in this
 case requires that one deals with a multi-temperature universe. We discuss the basic formalism
  which includes the multi-temperature nature of visible and hidden sectors in the analysis of
   phenomena observable in the visible sectors.  Specifically we discuss the application of the formalism
   for explaining the velocity dependence of dark matter cross sections as one extrapolates from galaxy scales
    to scales of galaxy clusters.  Here the dark photon exchange among dark fermions can
   produce the desired velocity dependent cross sections consistent with existing galactic cross section data
   indicating the existence of a new fifth (dark) force.
   We also discuss the possibility that the dark photon may constitute a significant portion of  dark matter.
    We demonstrate a realization of this possibility in a universe with two hidden sectors and
    with the visible sector and the hidden sectors
    in different heat baths which allows a satisfaction of the constraints
     that the dark photon have a lifetime larger than the age of the
    universe and that its relic density be consistent with Planck data. Future directions for further work
    are discussed.

\end{abstract}

\numberwithin{equation}{section}

\newpage

{ \hypersetup{colorlinks=black,linktocpage=true} \tableofcontents }

\section{Hidden sectors in supergravity and strings}

Most modern model building based on  supergravity and strings involve hidden sectors
which arise in a natural way in a variety of higher dimensional supergravity and string
compactifications.  Hidden sectors first arose in
the context of gravity mediated breaking of supersymmetry and were subsequently identified in
string model building (for a review see~\cite{Nath:2016qzm}).
 While the hidden sectors are neutral under the Standard Model (SM) gauge group
they can still  communicate with the visible sector via a variety of portals such as the Higgs portal~\cite{Patt:2006fw}
and  via kinetic mixing~\cite{Holdom:1985ag}
or  Stueckelberg mass mixings~\cite{Kors:2004dx}
 or both~\cite{Feldman:2006wb}
if the hidden sectors are charged under extra $U(1)$ gauge
groups. Further, the hidden sectors may contain matter~\cite{Kors:2004dx,Cheung:2007ut,Feldman:2006wb}.
 Because of feeble couplings between the hidden and the
  visible sectors,    the hidden sectors may reside in heat baths which are significantly colder than that of the
   visible sector. The feeble coupling between the hidden and the visible sectors would tend to
   thermalize the two sectors over time as long as the two sectors are coupled chemically and kinetically.
    However, such a thermalization in the early universe is not instantaneous and thus thermal evolution
    of the  coupled visible and hidden sectors must be taken into account in the study of the early universe and of the
    particle physics phenomena that carry signatures of physics of the early universe.

    A coupled evolution of the visible and the hidden sectors  brings in significant new     constraints~\cite{Aboubrahim:2020lnr,Aboubrahim:2021ycj}.
    First, the entropy conservation
    holds only  for  the sum of all entropies including that of the visible and of the hidden sectors.
    Some
    of the early works ignored this constraint and used separate entropy conservations for the
    visible and the hidden sectors which is invalid in the presence of coupling between the sectors.
    Second, the Friedman equations are modified since the energy density and the pressure density appearing there
     include those from the visible and from the hidden sectors which consequently also affect the evolution of the Hubble
     parameter.  Because of this, cosmology of the early universe
    is affected by the presence of hidden sectors.
    Since the visible sector and the hidden sectors may reside in different heat baths, a multi-temperature nature of
    the universe must be implemented for the proper understanding of the cosmological and particle physics
     phenomena. We address this issue in this note.

 \begin{figure}[H]
 \centering
   \includegraphics[width=0.5\textwidth]{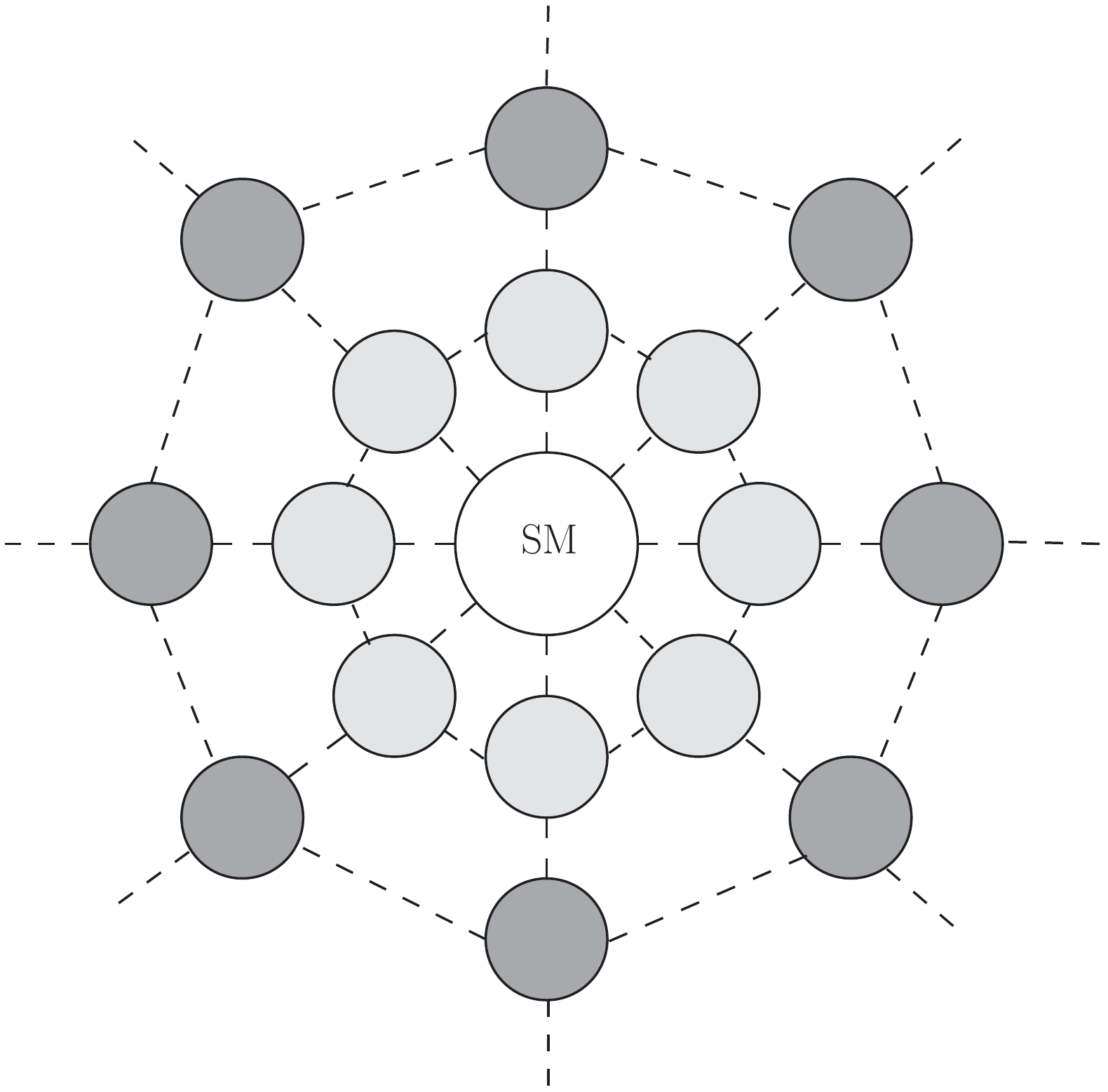}
   \caption{A schematic diagram exhibiting the SM extended by various hidden sectors,
   indicated by grey circles.
   There might be multiple hidden sectors beyond the SM
   that have either direct couplings or indirect couplings with the SM.
   In general these hidden sectors will evolve separately and live in their own heat baths.}
\label{Fig.GHS}
\end{figure}

         The outline of the rest of the paper is as follows: In section~\ref{sec:multiT} we discuss the basic cosmological
          constraints that govern coupled visible and hidden sectors. Here we assume an arbitrary number of
          hidden sectors coupled to the visible sector (see Fig.~\ref{Fig.GHS}). We assume that each of the hidden
          sectors possesses a $U(1)$ gauge group which has kinetic mixing and Stueckelberg mass mixing
            with the hypercharge field and with each other. In section~\ref{sec:hs} we discuss the dynamical equations that
            allow one to evolve the hidden sectors and the visible sector and compute  the temperatures
            of the hidden sectors as a  function of a reference temperature. Here we discuss two explicit
            examples. In the  first example,  we have one hidden sector and only one temperature ratio needs to be
            evolved, while in the second example we consider two hidden sectors and we have two temperature
             ratios that need to be evolved.
                In section~\ref{sec:darkforce}, we consider a hidden sector with a dark fermion and a dark photon where the dark fermion is the
             dark matter and the dark photon acts as a force mediator between the dark fermions. Assuming the mass of
             the mediator to be much smaller than the mass of the dark fermion, we fit the velocity dependent cross sections
             of dark fermionic dark matter from galaxy scales to scales of galaxy clusters.  In section~\ref{sec:darkphoton}, we discuss
              the possibility of a dark photon being dark matter. This possibility is difficult to realize with just one hidden sector,
               since the twin constraints that the dark photon relic density be consistent
              with the Planck data and the constraint that the lifetime of the dark photon be larger than the age of the
              universe are difficult to satisfy in this case. For this reason in this section we consider two hidden sectors and show
              that a dark photon as dark matter can be realized in this framework.
                     In section~\ref{sec:bbn} we discuss  the $\Delta N_{\text{eff}}$ constraint
              on the hidden sectors. In section~\ref{sec:future} we discuss future directions and conclusions in section~\ref{sec:conc}.

\section{Equations that govern a multi-temperature universe}
\label{sec:multiT}

If hidden sectors exist they contribute to the pressure and the energy density of the universe and enter in
the Friedman equations.
    Thus in the presence of
 several hidden sectors which in general are at different temperatures, the  Friedman equations read
  \begin{align}
 \label{FR1}
 H^2 &= \frac{8\pi G_N} {3} \left[\rho_v(T) + \sum_i \rho_i(T_i)\right],  \\
 \frac{ \ddot{a}}{a}&= - \frac{4\pi G_N}{3} \left[(\rho_v+3p_v)(T) +
  \sum_i (\rho_i +3 p_i)(T_i)\right],
 \label{FR2}
 \end{align}
 where $a$ is the scale factor in an FRW universe,  $H$ is the Hubble parameter,
 $G_N$ is Newton's gravitational constant, $p_v, \rho_v$
 are the pressure density and the energy density for the visible sector which are functions of the visible sector temperature
 $T$, and $p_i,\rho_i$ are the pressure density
 and the energy density for the  $i^{\rm th}$ hidden sector and are functions of the hidden sector temperature
 $T_i$.    Further, in the presence of couplings between the visible and the hidden sectors, it is only the
 total entropy $S$, where $S=a^3 s$ with $s$ being the entropy density and $a$ the scale factor, which is conserved, i.e.,
 $\rm{d}(sa^3)/\rm{d}t=0$ leads to
  \begin{align}
  & {\rm d} s/{\rm d} t + 3 Hs=0,\nonumber\\
&s=\frac{2\pi^2}{45} \left(h^v_{\rm eff}(T)  T^3+\sum_i h_{\rm eff}^{h_i} (T_i) T_i^3\right),
 \end{align}
 where $h^v_{\rm eff}(T)$ is the effective entropy degrees  of freedom for the visible sector at temperature $T$
 and $h_{\rm eff}^{h_i} (T_i)$
 is the effective entropy degrees of freedom for the $i^{\rm th}$ hidden sector at temperature $T_i$.
 Further, the total energy density satisfies the conservation equation,
  \begin{align}
  \frac{d\rho}{dt}+ 3H(\rho + p) =0.
  \label{FR3}
 \end{align}
 One may also exhibit the equations of motion satisfied by the individual energy density in each sector so that
   \begin{align}
  &\frac{d\rho_v}{dt}+ 3H(\rho_v + p_v) =j_v,\nonumber\\
  &\frac{d\rho_i}{dt}+ 3H(\rho_i + p_i) =j_i, (i=1\cdots n),
    \label{FRx}
 \end{align}
 where $j_v, j_i$ are the corresponding source functions which
 are determined by the couplings among the visible and the hidden sectors.
 They are discussed in detail in~\cite{Aboubrahim:2020lnr,Aboubrahim:2021ycj}.
   In order to study the correlated temperature dependence in the evolution equations it is then convenient to
  write the evolution equations in terms of temperature rather than time.
  The equation relating the two is given by
  \begin{align}
 \frac{dT}{dt}= - \frac{4\lambda\rho}{d\rho/dT} H.
 \label{FR4}
 \end{align}
 Here $\lambda=\frac{3}{4} (1+ p/\rho)$ and  where $\lambda=1$ is for the radiation dominated era and
 $\lambda=3/4$ for the matter dominated universe.  Since the coupled equations involve different temperatures,
the temperature evolution of the hidden sectors relative to the visible sectors is determined by the functions
\begin{align}
\xi_i= \frac{T_i}{T}, ~i=1\cdots n,
\end{align}
 which requires that one generate evolution equations for $\xi_i$ or equivalently for $\eta_i\equiv \xi_i^{-1}$.
 In deriving the evolution equations one needs to choose a temperature clock, which is a reference temperature,
in terms of which all other temperatures are measured. This reference temperature
  can be either the visible sector temperature or the temperature of one of the hidden sectors. Thus if $T$ is chosen
 as the clock then one needs $n$ number of evolution equations for the quantities $d\xi_i/dT~(i=1\cdots n)$.\\

 As indicated in Fig.~\ref{Fig.GHS}, in general the universe can have a large number of hidden sectors,
some of which have direct coupling with the SM while others have indirect or no coupling with the SM.
For the case of additional $U(1)$'s, each hidden sector may have matter fermions and gauge bosons.
The total Lagrangian  is given by
\begin{equation}
\mathcal{L} = \mathcal{L}_{\rm vis} + \mathcal{L}_{\rm hid} + \mathcal{L}_{\rm vh}\,,
\end{equation}
where  $\mathcal{L}_{\rm vis}$ is the SM Lagrangian, $\mathcal{L}_{\rm hid}$ is the Lagrangian for the
hidden sector and  $\mathcal{L}_{\rm vh}$ is the Lagrangian which involves the couplings between the  visible sector
and the hidden sector. In general the hidden sector involving $n$ number of hidden sectors and invariant under $n$ $U(1)$
gauge transformations is given by
\begin{equation}
\mathcal{L}_{\rm hid} = -\frac{1}{4} \sum_i F_{i \mu\nu}F^{\mu\nu}_i
- \frac{1}{i}\sum_{i}  \bar{\chi}_{i} \gamma_\mu (\partial^\mu -ig_i Q_iC^\mu_i)\chi_{i}
- \sum_{i} m_{i} \bar{\chi}_{i}{\chi}_{i} \,,
\end{equation}
where we sum over all hidden sectors, and $\chi_{i}$ indicates the $i$th fermion from the $i$th hidden sector with mass $m_{i}$.

Finally, $\mathcal{L}_{\rm vh}$ contains  interactions
between the visible sector and hidden sectors,
as well as interactions among the hidden sector gauge bosons via kinetic mixing and via Stueckelberg mass mixing
and is  given by
\begin{equation}
\mathcal{L}_{\rm vh} =
- \sum_{i}\frac{\delta_{i}}{2} B_{\mu\nu}F^{\mu\nu}_i
- \sum_{i<j}\frac{\delta^h_{ij}}{2} F_{i\,\mu\nu}F^{\mu\nu}_j
-  \frac{1}{2} \left(M_{B} B^\mu + \sum_i M_i C^\mu_i + \sum_i\partial_\mu \sigma_i\right)^2\,,
\label{st1}
\end{equation}
where $\delta_{i}$ is the kinetic mixing parameter between the $i$th hidden sector with the hypercharge,
$\delta^h_{ij}$ is the kinetic mixing parameter among the $i$th and the $j$th hidden sectors.
In Eq.~(\ref{st1}), $\sigma_i$ are axionic fields which have appropriate transformations under the $U(1)_Y$ and
$U(1)_{X_i}$ to guarantee the gauge invariance of Eq.~(\ref{st1}) and we drop them when working in the unitary gauge.
As noted earlier, in general both kinetic mixing and Stueckelberg mass mixing between the visible sector and the hidden  gauge fields
play a role in the dynamics involving the visible and the hidden sectors~\cite{Aboubrahim:2020lnr,Aboubrahim:2021ycj}.

\section{Evolution of hidden sector temperatures}
\label{sec:hs}

\begin{figure}
 \includegraphics[width=0.5\textwidth]{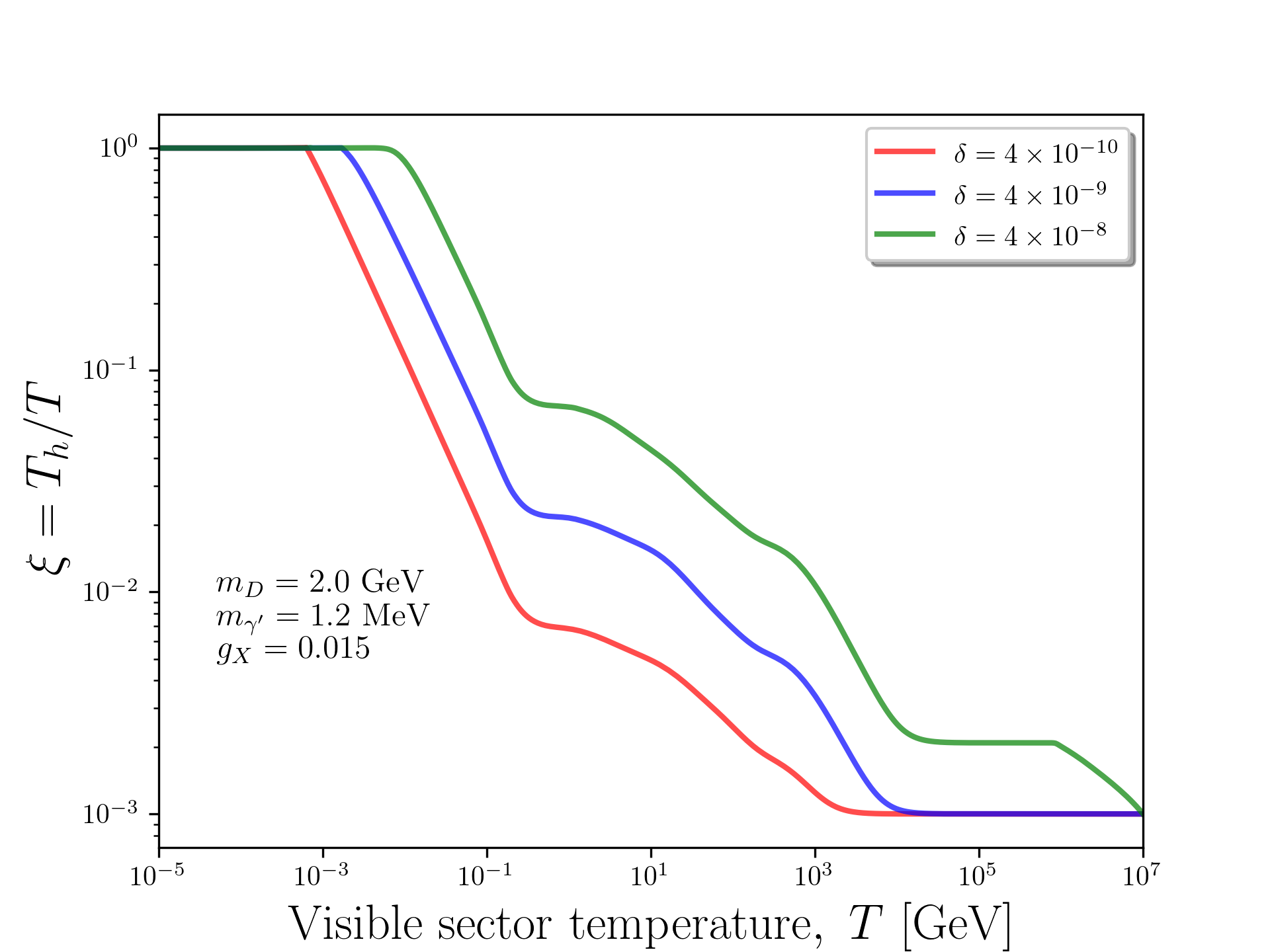}
 \includegraphics[width=0.48\textwidth]{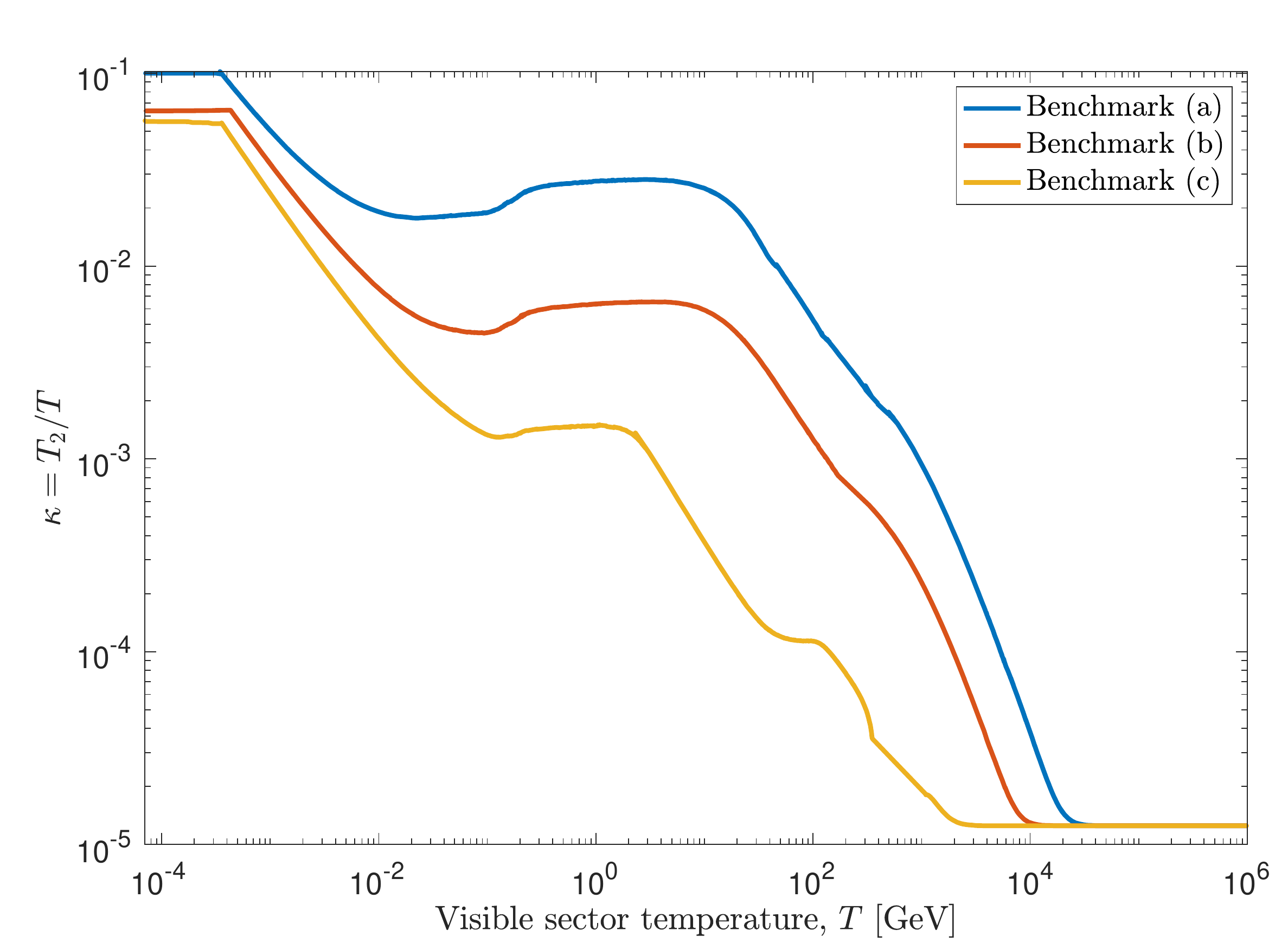}
\caption{Left panel:
Evolution of $\xi$ for the case of one hidden sector with the visible sector temperature for three values of the gauge kinetic mixing $\delta$ as given in the
 upper
 Table~\ref{tab1}.  One may note that the thermalization of the hidden and the visible sector is slower  more feeble
 the coupling between the sectors is.
 Right panel: Evolution of $\kappa=T_2/T_1$  as a function of the visible sector temperature $T$ for three benchmarks (a), (b) and (c)
 as given in the lower
 Table~\ref{tab1}. Figures from~\cite{Aboubrahim:2020lnr,Aboubrahim:2021ycj}.
 }
\label{fig1}
\end{figure}

As was noted above, the temperatures of the visible and the hidden sectors
could be different~\cite{Aboubrahim:2020lnr,Aboubrahim:2021ycj}
(for earlier literature see~\cite{Hambye:2019dwd,Foot:2014uba,Foot:2016wvj}).
We discuss now the explicit evolution of the temperatures in the hidden sector vs the visible sector for two
 cases: one for the case of a single  hidden sector and the other when there are two hidden sectors.
 First we discuss the case of a single hidden sector where  we choose for convenience our reference temperature to be
 the hidden sector temperature $T_h$ and a direct computation gives~\cite{Aboubrahim:2020lnr}
\begin{equation}
\frac{{\rm d} \eta}{{\rm d} T_h}= - \frac{A_v}{B_v} +  \left(\frac{\lambda \rho_v+ \rho_h( \lambda-\lambda_h)+
j_h/(4H)}{\lambda_h\rho_h- j_h/(4H)}\right)\frac{d\rho_h/dT_h}{B_v},
\label{y1}
\end{equation}
where
 \begin{align}
\label{y2}
A_v&=\frac{\pi^2}{30}\left(\frac{{\rm d} g_{\rm eff}^v}{{\rm d} T}\eta^5 T_h^4+4g_{\rm eff}^v\eta^4 T_h^3\right),\\
B_v&=\frac{\pi^2}{30}\left(\frac{{\rm d} g_{\rm eff}^v}{{\rm d} T}\eta^4 T_h^5+4g_{\rm eff}^v\eta^3 T_h^4\right).
\label{y3}
\end{align}
Here $\lambda$ is as defined after Eq.~(\ref{FR4})  and $\lambda_h= \dfrac{3}{4}\left(1+ \dfrac{p_h}{\rho_h}\right)$, where
 $\rho_h$ and $p_h$ are determined
in terms of the density of hidden sector particles. We implement this for the specific model given by the Lagrangian
 \begin{align}
 \mathcal{L}=& -\frac{1}{4} C^{\mu\nu} C_{\mu\nu} - g_X \bar D \gamma^\mu DC_\mu  -m_D \bar D D \nonumber\\
 &- \frac{\delta}{2} C^{\mu\nu} B_{\mu\nu} - \frac{1}{2}(M_1C_\mu + M_2 B_\mu + \partial_\mu \sigma)^2,
 \label{Lag1hs}
 \end{align}
 where $C_\mu$ is the gauge field of the hidden sector associated
 with the $U(1)_X$ gauge group,  $D$ is a Dirac fermion which is charged under $U(1)_X$ and $\delta$ is the kinetic mixing parameter. We assume that initially the hidden sector is colder than the
 visible sector but they begin to thermalize with time and eventually attain thermal equilibrium.
  However, as shown
 in the left panel of Fig.~\ref{fig1}, the rate at which they thermalize varies depending on the size of the couplings. As expected the
 thermalization is slower for smaller $\delta$ and becomes faster as $\delta$ increases.

 \begin{table}
 \begin{center}
\begin{tabular}{|cccccc|}
\multicolumn{6}{c}{Case of one hidden sector} \\
\hline
&Model & $m_D$ (GeV) & $M_1$ (MeV) & $g_X$ & $\delta ~(\times 10^{-9}$)\\
\hline
&(a) & 1.50 & 1.20 & 0.016 & $28$ \\
&(b) & 2.0 & 1.22 & 0.014 & $4.0$\\
&(c) & 2.16 & 1.13 & 0.015 & $4.7$\\
\hline
\end{tabular}
\end{center}
\centering
\begin{tabular}{|cccccccccc|}
\multicolumn{10}{c}{Case of two hidden sectors} \\
\hline
Model & $m_D$ & $M_1$ & $M_3$ & $M_4$ & $\delta_1$ & $\delta_2$ & $m_{Z'}$ & $m_{\gamma'}$ & $\Omega h^2$ \\
\hline
(a) & 1.00 & 4.50 & 0.0 & 0.43 & $4.0\times 10^{-10}$ & 0.40 & 4.90 & 0.43 & 0.124 \\
(b) & 0.50 & 4.50 & 0.0 & 0.47 & $6.5\times 10^{-11}$ & 0.40 & 4.90 & 0.47 & 0.103 \\
(c) & 0.05 & 4.50 & 0.0 & 0.45 & $5.6\times 10^{-12}$ & 0.40 & 4.91 & 0.45 & 0.102 \\
(d) & 0.62 & 4.50 & -5.0 & 0.45 & $4.0\times 10^{-10}$ & 0.05 & 6.76 & 0.30 & 0.108 \\
\hline
\end{tabular}
\caption{\label{tab1}
Upper table:
Benchmarks for the case when there is only one hidden sector.
Lower table:
Benchmarks for the case when there are two hidden sectors where
 $g_X=0.95$ and masses are in MeV except $m_D$ which is in GeV. Taken from~\cite{Aboubrahim:2020lnr,Aboubrahim:2021ycj}.}
\end{table}

\begin{figure}
\begin{center}
 \includegraphics[width=0.6\textwidth]{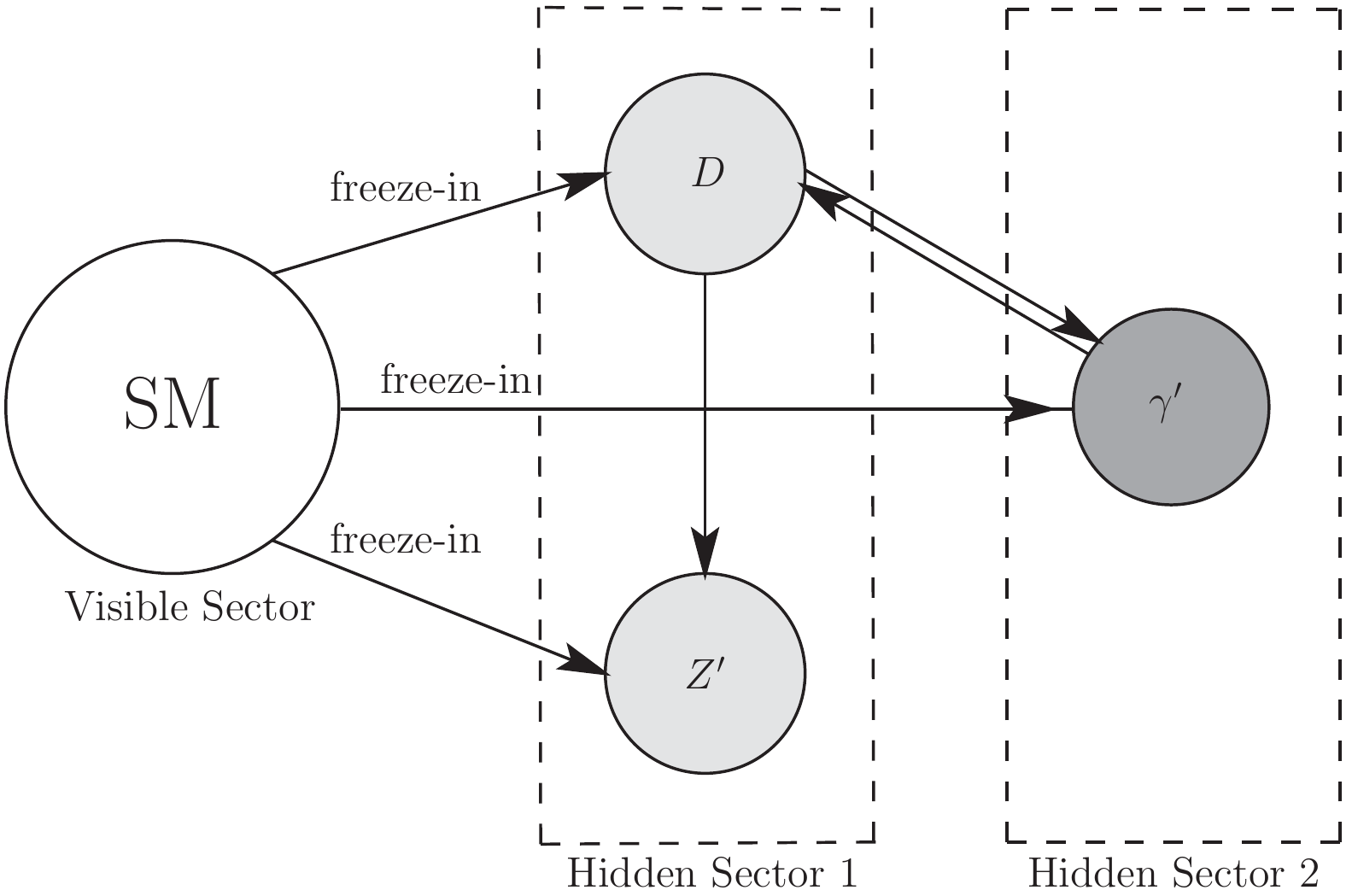}
\caption{Schematic diagram of the visible sector interacting with two hidden sectors. Hidden sector 1 has a dark $Z'$ and a dark fermion $D$ and has direct coupling with the visible sector. Hidden sector 2 has a dark photon $\gamma^\prime$ but no dark fermion and
direct coupling with hidden sector 1 and only indirect coupling with the visible sector via hidden sector 1.}
\label{fig.hid2}
\end{center}
\end{figure}

Next we consider the case of two hidden sectors which together have an extended gauge invariance $U(1)_{X_1}\times
U(1)_{X_2}$ where hidden sector 1 with gauge invariance $U(1)_{X_1}$ has the gauge field $C^\mu$ and a Dirac fermion $D$
charged under $U(1)_{X_1}$ and has kinetic coupling $\delta_1$ to the visible sector.
The hidden sector 2 with gauge invariance $U(1)_{X_2}$ has a gauge field $D^\mu$ and  a kinetic coupling $\delta_2$  and a  Stueckelberg mass mixing via the mass
parameter $M_3$ with the hidden sector 1. However, it has no direct coupling
with the visible sector, see Eq.~(\ref{Ltot}) and  Fig.~\ref{fig.hid2}. Thus the Lagrangian
involving the fields of the hidden sectors is given by
\begin{align}
 \mathcal{L}_{\text{h}}+ \mathcal{L}_{\text{hv}}
 =&-\frac{1}{4}C^{\mu\nu}C_{\mu\nu}-\frac{1}{4}D^{\mu\nu}D_{\mu\nu}-\frac{\delta_1}{2}B^{\mu\nu}C_{\mu\nu}-\frac{\delta_2}{2}C^{\mu\nu}D_{\mu\nu}\nonumber\\
&-\frac{1}{2}(M_1 C_{\mu})^2-\frac{1}{2}(M_3 C_{\mu}+M_4D_{\mu})^2-m_D \bar{D}D
+ g_X \bar D\gamma^{\mu}DC_{\mu}.
\label{Ltot}
\end{align}
The total Lagrangian $\mathcal{L}=\mathcal{L}_{\text{v}}+\mathcal{L}_{\text{h}}+ \mathcal{L}_{\text{hv}}$
must be canonically normalized and its mass eigenstates
in the neutral vector boson sector consist of
$\gamma', Z', Z, \gamma$. Here $Z$ and $\gamma$ reside in the visible sector, $Z'$ is the neutral massive gauge boson in
 hidden sector 1 and $\gamma'$ is  the dark photon which resides
in hidden sector 2. In this case we   evolve
temperatures $T_1$ of  hidden sector 1 and $T_2$ of hidden sector 2, and $T$ for the visible sector in a correlated
fashion. We define two functions that allow us to accomplish this
\begin{align}
\eta= \frac{T}{T_1}~~\text{and}~~\zeta= \frac{T_1}{T_2}.
\end{align}
It is then straightforward to deduce the following evolution equations for $\eta$ and $\zeta$~\cite{Aboubrahim:2021ycj}
\begin{align}
\label{boltz-4}
\frac{d\eta}{dT_1}=&-\frac{\eta}{T_1}+\left(\frac{4H\rho_v+j_1+j_2}{4H\rho_1-j_1}\right)\frac{d\rho_1/dT_1}{T_1\frac{d\rho_v}{dT}},\\
\frac{d\zeta}{dT_1}=&-\frac{\zeta}{T_1}+\left(\frac{4H\rho_2-j_2}{4H\rho_1-j_1}\right)\frac{d\rho_1/dT_1}{T_1\frac{d\rho_2}{dT_2}},
\label{boltz-5}
\end{align}
where we used $T_1$ as the clock.
Solution to these equations allows us to compute the temperatures in the hidden sectors 1 and 2
in terms of the visible sector temperature. To exhibit this numerically we consider a
 set of benchmarks as given in the lower
   Table~\ref{tab1}.  In this case the hidden sector 1 evolves  similar to the left panel of Fig.~\ref{fig1}
and the hidden sector 1 and the visible sector thermalize.
 However, the temperature in the hidden sector 2 evolves more slowly and the hidden sector 2 does not
 fully thermalize with the visible sector as exhibited in right panel of Fig.~\ref{fig1}.
  Here we end up with the ratio $T_2/T \leq 0.1$ for the model points exhibited. This ratio has direct implications for $\Delta N_{\rm eff}$ at BBN time as will be discussed
   later.

\section{A dark force and galaxy structure anomalies}
\label{sec:darkforce}

In this section we discuss how hidden sectors can affect cosmology in significant ways.
Here one area where hidden sectors could have direct impact is dark matter.
Thus currently the standard  $\Lambda$CDM is seen to work well at large scales. However, at galaxy scales
there seem to appear some anomalies. Often these are described as the missing satellite, cusp-core and the too-big-to-fail anomalies. It is possible they could be resolved by more comprehensive analyses using complex dynamics
and baryons along with CDM~\cite{Governato:2012fa}. However, other alternatives have also been proposed
such as self-interacting dark matter~\cite{Spergel:1999mh} which has attracted
attention~\cite{Tulin:2017ara,Vogelsberger:2012ku,Rocha:2012jg,Peter:2012jh,Zavala:2012us,Elbert:2014bma,Vogelsberger:2014pda,Fry:2015rta,Dooley:2016ajo,Buckley:2009in,Loeb:2010gj,Tulin:2012wi,Tulin:2013teo,Schutz:2014nka,Bringmann:2016din}.
Thus in this picture dark matter is self-interacting and collisional at galaxy scales, but collision-less at the scale of
galaxy clusters. Currently there exist data on dark matter cross sections over the range of scales
mentioned above~\cite{Robertson:2018anx,Sagunski:2020spe, Andrade:2020lqq,Elbert:2016dbb}
and we will refer to it as `data  from Dwarf Galaxy scales to
galaxy Clusters (DGC)'. The data indicates $\sigma/m=O(1) {\rm cm}^2/{\rm g}$ at small galaxy scales
and appears to exhibit a velocity dependence of
dark matter cross sections. Such a velocity dependence does not arise from conventional WIMP scattering.
In previous works, Yukawa potentials have been used to fit the data~\cite{Tulin:2017ara,Sagunski:2020spe}.
However, within this setup it is difficult to compute the relic density of dark matter.

 We discuss now a hidden sector model to accomplish that, i.e., satisfy the relic density and give a fit to DGC. The model we use
 is the first one discussed in section 3 (see Eq.~\ref{Lag1hs})),
 where we have one hidden sector with a dark fermion and a dark photon. In the analysis
 we assume that dark matter is constituted of dark fermions ($D$ and $\bar D$) and their self-interactions are generated via
 the exchange of dark photons. The dark photons, however, are unstable and decay before BBN and do not contribute to the
 relic density.  In the analysis we assume that the dark particles are produced by the freeze-in mechanism~\cite{Hall:2009bx,Aboubrahim:2019kpb,Aboubrahim:2020wah,Koren:2019iuv,Du:2020avz}.
  Here  even with feeble couplings to the visible sector
it is possible to generate $D$ and $\bar D$ in sufficient amounts
to saturate the dark matter relic
 density~\cite{Aboubrahim:2020lnr}.

\begin{figure}
 \begin{center}
   \includegraphics[width=0.6\textwidth]{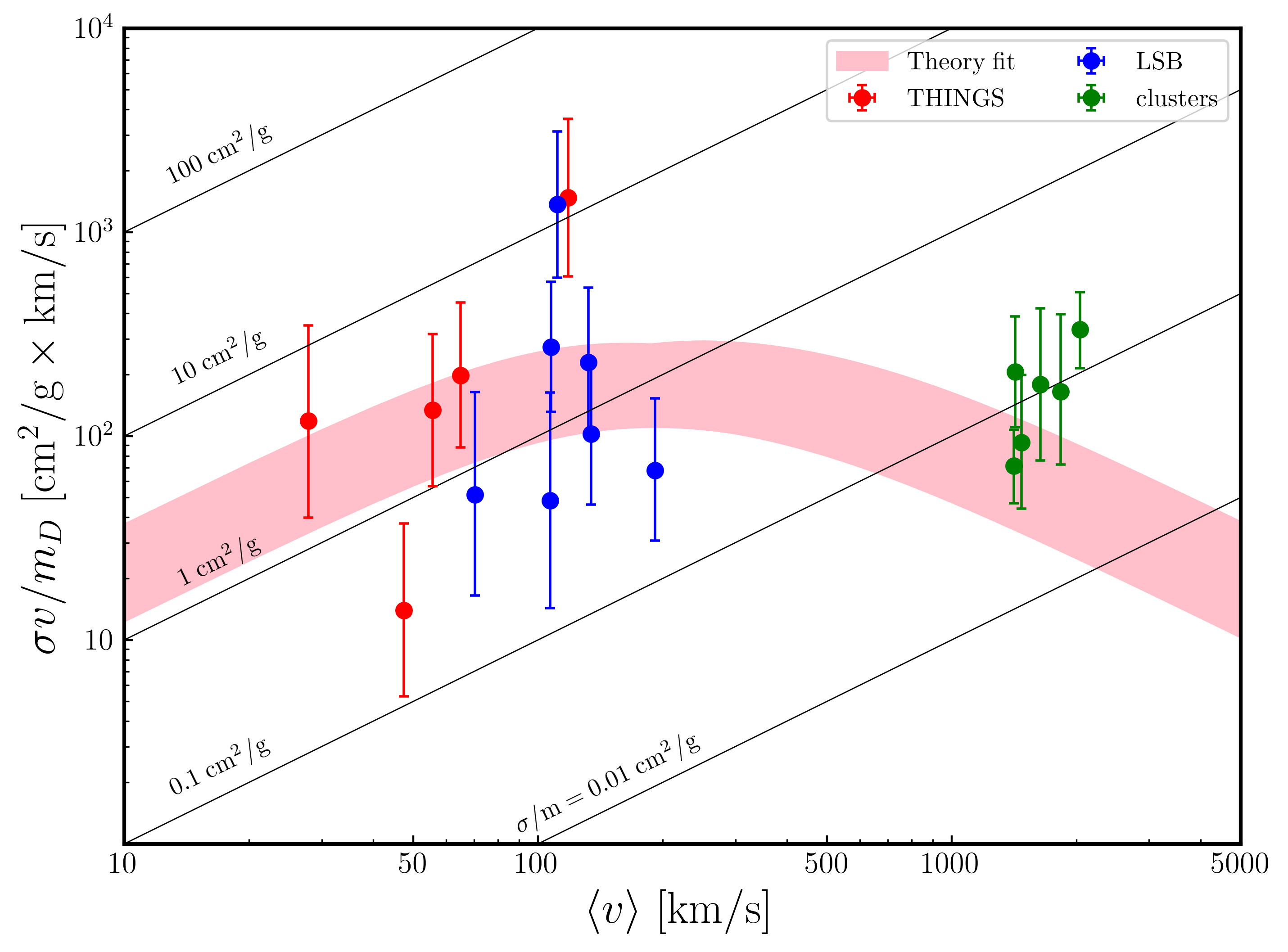}
   \caption{A plot of the theory prediction of
    $\sigma v/m_D$  versus  $\langle v\rangle$ (the average relative velocity of the DM particles
    in the halo) based on the model with one hidden sector with dark fermions self-interacting via exchange of dark photons with
    $m_{\gamma'} << m_D$,
     where the band represents the range of model points presented in~\cite{Aboubrahim:2020lnr}
     and the  DGC data is taken from the analysis of~\cite{Kaplinghat:2015aga,Sagunski:2020spe}.}
     	\label{sigma}
	\end{center}
\end{figure}

To fit the galactic cross section data, we compute  the self-interaction cross-section
\begin{equation}
\frac{d\sigma}{d\Omega}=\sum_{i=1}^3 \frac{\overline{|\mathcal{M}_i|^2}}{64\pi^2s},
\end{equation}
where $i=1\cdots 3$ sums over the three processes $D\bar D\to D\bar D$,  $DD\to DD$, and $\bar D \bar D\to \bar D \bar D$ and using the Lagrangian of Eq.~(\ref{Lag1hs}), one can compute $\sigma v/m_D$ where $v$ is the Moller velocity.
The explicit forms of $\mathcal{M}_i$ are given in~\cite{Aboubrahim:2020lnr}.
 A set of benchmarks which satisfy all of the current dark matter constraints are given in  upper Table~\ref{tab1}.
   A plot of $\sigma v/m_D$ is given in Fig.~\ref{sigma} and shows that a fit to DGC occurs for $\langle v\rangle$ from
   (10$-$3000) km/s.

\section{Dark photon as dark matter}
\label{sec:darkphoton}
\begin{figure}[H]
\centering
\includegraphics[width=0.8\textwidth]{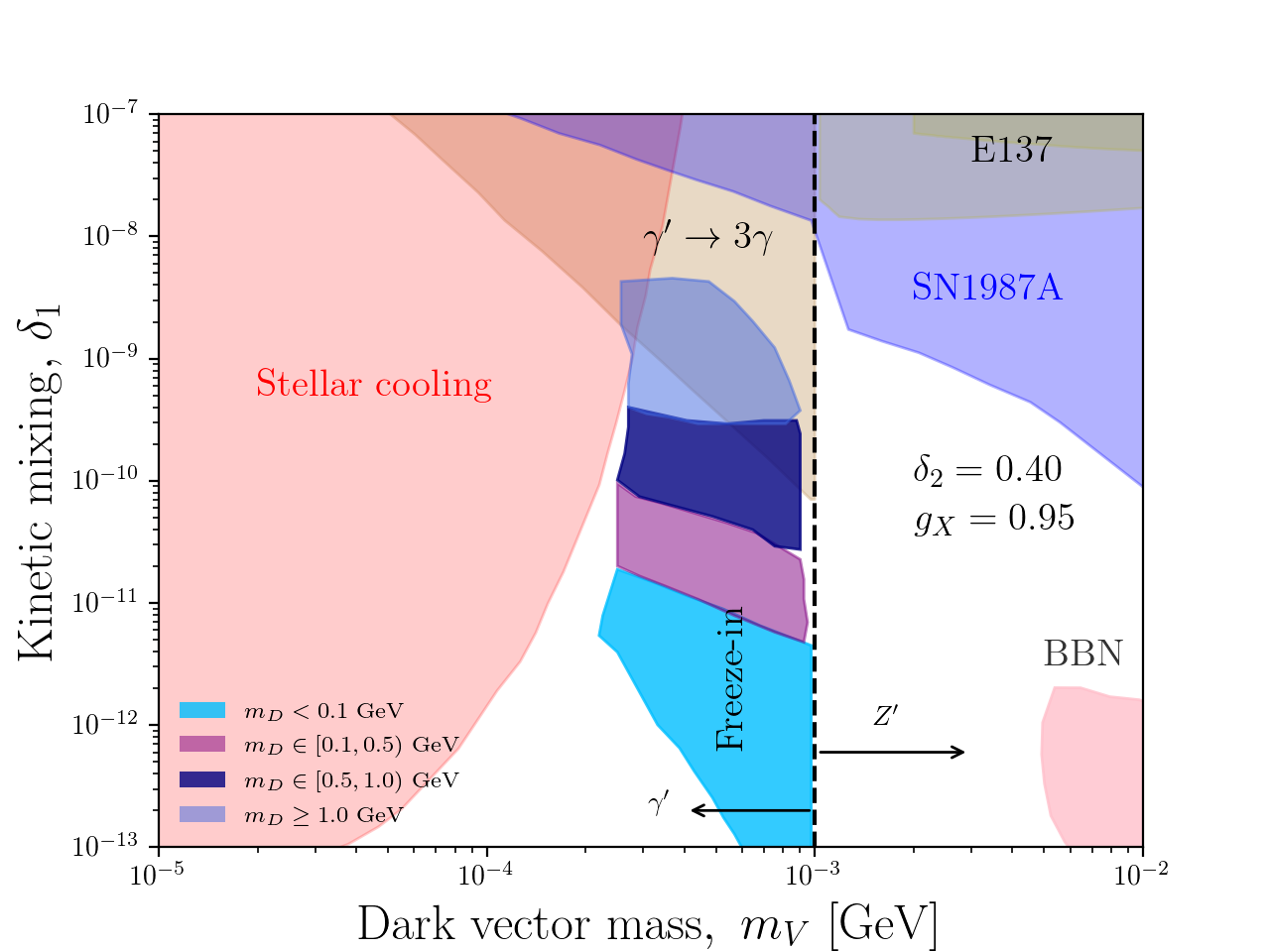}
\caption{An exhibition of the allowed region in the $\delta_1$ vs $m_{\gamma'}$ parameter space where a dark photon
is long lived with a lifetime larger than the age of the universe and can contribute a substantial portion of the
dark matter relic density. Here $\delta_2=0.4$ and  $m_{Z'}=(5-20)m_{\gamma'}$.
The regions where a dark photon can exist as dark matter are indicated by four colored vertical regions  corresponding to
the four ranges of $m_D$ indicated in the legend on the lower left hand corner.
Taken from~\cite{Aboubrahim:2021ycj}.
 The description of the various experimental constraints on the model are discussed in the text.}
\label{dm}
\end{figure}

As discussed above it is likely that dark matter resides partially or wholly in the hidden sector (for early work
see, e.g.,~\cite{Feldman:2006wd}).
 Here we discuss the
 possibility that the dark photon constitutes  a significant portion of dark matter.
 In order for this to occur, two important constraints need to be satisfied.
 First, the dark photon must have a lifetime larger than the age of the universe and second,  it must
 be produced copiously enough to constitute a significant portion of the dark matter relic density
 consistent with the Planck data~\cite{Aghanim:2018eyx}, i.e.,
\begin{equation}
\Omega h^2=0.1198\pm 0.0012.
\label{relic}
\end{equation}
It appears difficult to satisfy both these constraints with one hidden sector. This is so because for
the dark photon to be long lived the kinetic mixing needs to be extremely small. On the other hand to generate
a significant amount of relic density we need the kinetic mixing larger than what is required to obtain a long lived dark photon.  This tension
between the lifetime and the relic density  constraints is relaxed by invoking two hidden sectors, see Eq.~(\ref{Ltot}).
We note in passing
that the case of two hidden sectors was discussed in a different context in~\cite{Aboubrahim:2020afx}.
 In this case we have two kinetic mixings $\delta_1$ and $\delta_2$, and a mass mixing $M_3$
 which allows for a relaxation of the tension between the lifetime and the relic density constraints.
  In Fig.~\ref{dm} we exhibit the region of the parameter space where a dark photon with a lifetime
  greater than the age of the universe and a relic density  consistent with Eq.~(\ref{relic}) can be
  achieved. The excluded regions of the parameter arise from constraints from  experiments which
  include BaBar~\cite{Lees:2014xha}, CHARM~\cite{Bergsma:1985qz,Tsai:2019mtm}, NA48~\cite{Batley:2015lha},
  NA64~\cite{Banerjee:2018vgk,Banerjee:2019hmi}, E141~\cite{Riordan:1987aw} and $\nu$-CAL~\cite{Blumlein:1990ay,Blumlein:1991xh,Tsai:2019mtm}, electron and muon $g-2$~\cite{Endo:2012hp}, and E137~\cite{Andreas:2012mt,Bjorken:2009mm}. For a dark photon mass less than 1 MeV the strongest constraints
  arise from the decay $\gamma'\to 3 \gamma$~\cite{Essig:2013goa,Redondo:2008ec},
  from Supernova SN1987A~\cite{Chang:2016ntp},  and from stellar cooling~\cite{An:2013yfc}. The regions excluded
  by the BBN constraint are in pink. In the analysis we have used the code {darkcast}~\cite{Ilten:2018crw} to obtain most of the above mentioned limits.
  We note that in addition to the mechanism for production of the dark photon in the early universe discussed
  in the analysis here, one also has pure gravitational production of dark photons~\cite{Graham:2015rva,Ema:2019yrd,Ahmed:2020fhc}.  The gravitational production is dependent
  on the dark photon mass and on the reheat temperature. For the parameters in the analysis of  Fig.~\ref{dm} with a dark photon mass of $\sim 1$ MeV,  gravitational production will be
  suppressed if the reheat temperature lies below  $10^{11}$ GeV.

Regarding the lifetime of the dark photon there are two main decay channels: $\gamma'\to 3 \gamma$ and
$\gamma'\to \nu\bar \nu$.
The decay width of $\gamma'\to 3\gamma$ is given by~\cite{Pospelov:2008jk,McDermott:2017qcg}
\begin{align}
\Gamma_{\gamma^{\prime}\to 3\gamma}& =\frac{17\alpha^{3}\alpha^{\prime}}{2^{7}3^{6}5^{3}\pi^{3}}\frac{m_{\gamma^{\prime}}^{9}}{m_{e}^{8}}\approx4.70\times10^{-8}\alpha^{3}\alpha^{\prime}\frac{m_{\gamma^{\prime}}^{9}}{m_{e}^{8}}.
\end{align}
Here $\alpha=e^{2}/4\pi$, $\alpha^{\prime}=(k e)^{2}/4\pi$, $k=-\delta_{1}(\delta_{2}-\sin\beta)\cos\theta_w$,
where $\theta_w$ is the weak angle, and $\beta$ is defined such that $\tan\beta= M_3/M_4$ where $M_3$ and $M_4$ are
as given in Eq.~(\ref{Ltot}).
 For the parameters of model (d) at the bottom of
Table~\ref{tab1}, the decay lifetime
is $\tau_{\gamma'\to 3\gamma}\sim 5.3\times 10^{15}$ yrs. For $\gamma'\to \nu\bar \nu$ the decay width is
 given by
\begin{align}
\Gamma_{\gamma'\to \nu\bar{\nu}}&=\frac{g_2^2\delta_1^2(\delta_2-\sin\beta)^2\epsilon_{\gamma'}^4}{8\pi}m_{\gamma'}\tan^2\theta_w\,,
\end{align}
where $\epsilon_{\gamma'}= m_{\gamma'}/m_Z$.
In this case the dark photon lifetime for the illustrative model (d) is
$\tau_{\gamma'\to\nu\bar\nu}\sim 8.4\times 10^{21}$ yrs. Thus the dominant decays of the dark photon have
lifetime much larger than the age of the universe. Also as  can be seen from Table~\ref{tab1} one can generate
a relic density consistent with the Planck data.

\section{Hidden sectors and the BBN constraint}
\label{sec:bbn}

Hidden sectors bring in new degrees of freedom while BBN constraints strongly limit the extra neutrino
degrees of freedom. Thus the SM predicts $N_{\rm eff}=3.046$ at BBN time while the analysis
using data from BBN, CMB and BAO~\cite{Aghanim:2018eyx}
 gives a deviation from the SM of $\Delta N_{\rm eff}= 0.22$.
 Theoretically, if there are extra $\Delta n_b$ relativistic bosonic degrees of freedom beyond those of the SM
 they  contribute an extra amount to $\Delta N_{\rm eff}$ so that
  \begin{align}
 \Delta N_{\rm eff}\simeq \frac{4 \Delta n_b}{7}
 \left(\frac{11}{4}\right)^{\frac{4}{3}} \left(\frac{T_h}{T_{\gamma}}\right)^4,
     \end{align}
where $T_h$ is the hidden sector temperature where the new degrees of freedom reside. Thus it is clear that if the hidden sector is
very feebly coupled to the visible sector, it is likely that it may be far from thermal equilibrium with the visible sector at the BBN time.
Thus suppose $T_h/T_\gamma=0.2$. In this case the constraint $\Delta N_{\rm eff}=0.22$ implies that
$\Delta n_b$ could be as large as about 60. While this is an extreme example, it illustrates the importance of
evolving in an accurate fashion the temperature of the hidden sector temperature relative to that of the visible sector.

\section{Future directions}
\label{sec:future}

Hidden sectors and their implications for particle physics and for cosmology can be tested in a number of current
and future experiments which we discuss below. Thus
hidden sector dark matter, specifically light dark matter in the MeV$-$GeV range and below,  would be explored
in a number of dark matter detectors both via  direct detection and via indirect detection.
An example of this is the Xenon-1T detector which is sensitive to dark matter in the sub-GeV mass range and provides a testing ground for
specific models of hidden sector with dark fermions down scattering from bound electrons in xenon in an inelastic
 exothermic process~\cite{Aprile:2020tmw}
  (for theoretical analyses of these see~\cite{Aboubrahim:2020iwb} and the references therein). Further improvements in experiment with Xenon-nT
 will lead to  greater sensitivity to hidden sector dark matter.
PandaX-II liquid xenon detector~\cite{Cheng:2021fqb} also measures  scattering cross section between the dark matter and electrons and has placed the most stringent limits in the dark matter range of 15$-$30 MeV and a corresponding
DM-electron cross section from $2\times 10^{-37} -3.1\times 10^{-38}$ cm$^2$. 
PandaX-II also explored constraints specifically for self-interacting dark matter with mass range of 1$-$100~GeV with a light mediator mass from 1~MeV to 1~GeV~\cite{Yang:2021adi}.
Further, much smaller mass ranges for the hidden sector dark matter will be explored by future dark matter detectors. Thus
dark matter-electron scattering from aromatic organic targets~\cite{Blanco:2019lrf,Blanco:2021hlm}
presents a new area of dark matter detection for light dark matter.
Here scattering events involve  dark matter interactions with electrons which reside in molecular orbitals
which have excitation energies of a few eV.  Such detectors are sensitive to dark matter which are as light as a few MeV.

Another dark matter experiment suitable for hidden sector dark matter is CDEX-10~\cite{She:2019skm} at the China Jinping Underground Laboratory which uses a germanium detector and has already
 put direct detection constraints on dark photons. Thus it
 sets stringent limits on the kinetic mixing parameter $\delta$ in the mass range of 0.1$-$10 KeV.
The superCDMS experiment~\cite{Amaral:2020ryn}  constraints the dark photon kinetic mixing parameter $\delta$ in the mass range  (1.2$-$50) eV.
Hidden sector dark matter can also  be detected via indirect means. Thus, for example, hidden sector dark fermions $\chi$
can annihilate via the process $\bar \chi \chi\to \gamma \gamma$. For light hidden sector dark matter in the  sub-GeV
range, the $X$-ray and $\gamma$-ray spectrum will be typically soft which, however, can be detectable by
e-ASTROGAM~\cite{DeAngelis:2016slk}  which is an observatory space mission dedicated to the study in the photon energy range from 0.3 MeV to 3 GeV~\cite{DeAngelis:2016slk} which is an energy range with sensitivity not accessible to other $X$-ray and $\gamma$-ray searches such as Fermi-LAT, EGRET, COMPTEL and SPI.

    As noted, models of hidden sector dark matter are very sensitive to the relative temperature of the hidden sector
    in which they reside vs that of the visible sector. For this reason accurate thermal evolution of the hidden sector and of the visible
    sector heat baths are essential. The formalism discussed here allows one to accomplish that and thus relevant
    for further study as analyses
    along these lines would be needed to test hidden sector models of the dark photon and of other
    astrophysical phenomena with data in the future.

     \section{Conclusion}
     \label{sec:conc}

     Modern theories of particle physics and cosmology based on supergravity and strings necessarily involve
     hidden sectors needed for breaking of supersymmetry. However, the hidden sectors needed in supersymmetry
     breaking are not the only ones that appear in supergravity and string compactifications. In general there
     can be additional hidden sectors many of them endowed with gauge groups. While neutral under the SM
     gauge groups, the hidden sector can communicate with the visible sector in a variety of ways such as
      via Higgs portals and via kinetic mixing and Stueckelberg mass mixing between the hidden sectors and the visible
      sector if the hidden sectors posses $U(1)$ gauge symmetries. In this work we point out that an accurate
       treatment of a visible sector coupled to hidden sectors leads to a multi-temperature universe with the visible sector
       and the hidden sector residing in different heat baths.  A proper treatment of the coupled system then
       involves correlated evolution of the temperatures for the coupled sectors. In this work we have discussed the
       general formalism for accomplishing that.
         We also discussed two specific cases of multi-temperature
       evolution where the first one involves one hidden sector and the second two hidden sectors.
        For the first case one needs only one evolution parameter for the simultaneous evolution of the two temperatures,
        i.e., $\xi= T_1/T$ where $T_1$ is the temperature of the hidden sector and $T$ is the temperature of the
        visible sector. For the case of two hidden sectors we need in addition the evolution of the parameter $\zeta=T_2/T_1$.
        The evolution equations for $\xi$ and $\zeta$ were discussed in section~\ref{sec:hs}.

        Applications of the formalism were given in sections~\ref{sec:darkforce} and~\ref{sec:darkphoton}. In section~\ref{sec:darkforce}, we discuss a model based on one
             hidden sector where the hidden sector contains a dark fermion which acts as dark matter  and the dark
             photon which mediates the interaction between the dark fermions. This model leads to self-interacting dark
             matter which has a velocity dependent cross section and allows one to fit the galactic data from
             galaxy scales where the dark matter appears to behaves collisional to the
     scale of galaxy clusters where the data indicates that dark matter is collision-less. In section~\ref{sec:darkphoton} we discussed the
     possibility that the dark photon constitutes the bulk of dark matter in the universe. Here it is shown  that
     to generate the desired relic density and also satisfy the constraint that the dark photon lifetime exceeds the
     lifetime of the universe, one needs two hidden sectors. It is shown in this case that while the hidden sector
     thermalizes with the visible sector at BBN time, this is not the case for hidden sector 2 for which the ratio $T_2/T$
     is smaller than unity and is helpful in the satisfaction of the BBN constraint on the degrees of freedom.
       The formalism discussed here is helpful in building realistic particle physics which include  dark particles and dark forces
       as well as in cosmology studies which include hidden sectors in the analysis.  \\~\\

\textbf{Acknowledgments: }
The research of AA was supported by the BMBF under contract 05H18PMCC1.
 WZF was supported in part by the National Natural Science Foundation of China under Grant No. 11905158 and No. 11935009. The research of PN and ZW was supported in part by the NSF Grant PHY-1913328.

\end{document}